\documentclass[twocolumn,english,aps]{revtex4}
\usepackage[T1]{fontenc}
\usepackage[latin9]{inputenc}
\usepackage{amsmath}
\usepackage{amssymb}
\usepackage{graphicx}
\usepackage{esint}



\usepackage{babel}
\begin{document}

\title{Natural Ordering and Uniform Tidal Effects in Globally Coupled Phase
Oscillator Models}

\pacs{05.45.Xt 64.60.an}

\author{David Mertens}

\affiliation{Department of Physics and Astronomy, Dickinson College, Carlisle,
Pennsylvania 17013, USA}
\begin{abstract}
Globally coupled phase oscillator models, such as the Kuramoto model,
exhibit spontaneous collective synchronization. Such models can be
restated in terms of interactions within and between subsets of oscillators.
An approximation for the internal structure of coherent subsets of
oscillators can be made based on the observations of natural ordering
and uniform tidal effects. The approximation is seen to perform well
for predicting the microstructure in a variety of phase oscillator
models.
\end{abstract}
\maketitle
Spontaneous collective synchronization among nearly identical phase
oscillators is a classic emergent phenomenon and phase transition
\cite{strogatz2000fromkuramoto}. The first theoretical model to provide
insight into this transition was provided by Winfree in 1967 \cite{winfree1967biological}
and takes the form
\begin{equation}
\dot{\theta}_{i}=\omega_{i}+\frac{K}{N}\sum_{j=1}^{N}P\left(\theta_{j}\right)Q\left(\theta_{i}\right).\label{eq:Winfree-model}
\end{equation}
The functions $P$ and $Q$ must be $2\pi$-periodic, and so can be
expressed as trigonometric functions of angular differences and sums.
Kuramoto considered a simpler coupling \cite{kuramoto1975selfentrainment}
given by:
\begin{equation}
\dot{\theta}_{i}=\omega_{i}+\frac{K}{N}\sum_{j=1}^{N}\Gamma\left(\theta_{j}-\theta_{i}\right).\label{eq:Kuramoto-model-full-sum}
\end{equation}
The simplest coupling is $\Gamma\left(\Delta\theta\right)=\sin\left(\Delta\theta\right)$.
Such a coupling tends to draw two nearby oscillators together, while
the disorder in the natural rotation rates, $\omega_{i}$, tends to
spread them out. For many choices of $P$ and $Q$ or of $\Gamma$,
if the coupling $K$ is strong enough and the variation of the $\omega_{i}$
is small enough, a subset of the population will mutually entrain:
they will lock into relative positions that are essentially fixed
after an initial transient. Winfree argued that the fraction of mutually
entrained oscillators exhibits a phase transition in the coupling
$K$, and Kuramoto analytically calculated the magnitude of the entrained
fraction for his model. Various models have been considered by other
authors \cite{ariaratnam2001phasediagram,sakaguchi1986asoluble} and
together they have been used to describe the collective behavior of
many physical systems including coupled Josephson junctions \cite{weisenfeld1998frequency},
laser arrays \cite{kourtchatov1995theoryof}, electronic auto-oscillators
\cite{temirbayev2012experiments}, electrochemical oscillators \cite{kiss2002emerging,kiss2008resonance},
and acoustically coupled mechanical rotors \cite{mertens2011synchronization,mertens2011individual}.

Globally coupled phase oscillator models can be rewritten exactly
as mean field models. Because of this, most research has focused on
approximating the behavior of the mean field. The most important results
are found with the approximation of large system size, $N\to\infty$
\cite{acebron2005thekuramoto,strogatz2000fromkuramoto,ott2008lowdimensional}.
Work focusing on finite populations relies on ensemble averaging to
make any meaningfully general claims \cite{daido1989intrinsic,chulho2013extended,buice2007correlations}.
Unfortunately, the only published mechanism for elucidating population-specific
tipping points is to measure or simulate the behavior directly.

\paragraph{Presence of coherent subsets}

\begin{figure}
\includegraphics[width=0.8\columnwidth]{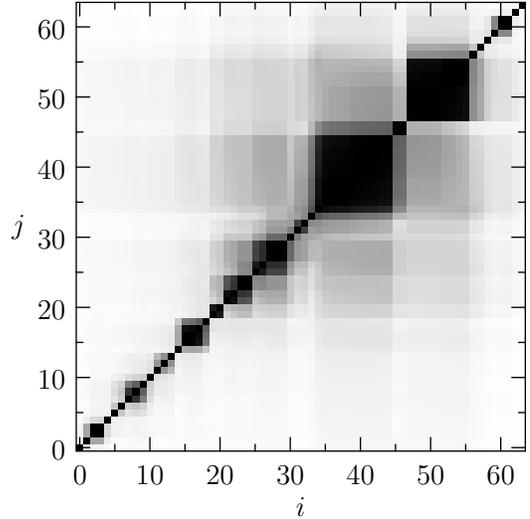}

\protect\caption{\label{fig:correlation-matrix}Pairwise correlations between all oscillators
in a simulation of the Kuramoto model, $N=64$. Lighter and darker
intensity indicates smaller and larger values of $\rho_{ij}$, respectively.
The indices were not optimized for block diagonal structure but were
simply sorted according to $\omega_{i}$. For this simulation, the
coupling strength is equal to the width of the population's Gaussian
distribution, placing it well below the critical coupling.}
\end{figure}

The mean field reformulation is exact, but it theoretically obscures
an important observation. Figure \ref{fig:correlation-matrix} illustrates
how groups of oscillators clump into groups (such as explored by Maistrenko
et al. \cite{maistrenko2004mechanism}). The intensity indicates the
value of $\rho_{ij}$, defined as
\begin{equation}
\rho_{ij}e^{\imath\Delta_{ij}}\equiv\frac{1}{T}\int_{0}^{T}e^{\imath\left(\theta_{i}-\theta_{j}\right)}dt,
\end{equation}
using $\imath\equiv\sqrt{-1}$. The obvious block diagonal structure
is obtained simply by sorting the indices according to natural frequency
$\omega_{i}$. Subsets of oscillators with nearby natural frequencies
mutually entrain. Although not indicated in this figure, oscillators
in these clumps also appear to act in concert, suggesting that their
degrees of freedom are not independent. Taken together, these observations
indicate that finite-sized phase oscillator models exhibit rich internal
structure.

The presence of coherent subsets gives rise to a number of interesting
questions. How do we best quantify the behavior of these coherent
subsets? What approximations must be made to express the dynamics
of the model in terms of the subsets' dynamics? How do the subsets
evolve? Can we approximate the structure or evolution of the subsets?
All of these questions will be addressed in this letter.

\paragraph{Subset reformulation}

The first papers on synchronization established that globally coupled
phase oscillator models can be written exactly in terms of mean fields
\cite{winfree1967biological,kuramoto1975selfentrainment}. For models
with coupling that depends only on the first harmonic, such as the
Kuramoto model, we use the mean field
\begin{equation}
r\,e^{\imath\psi}\equiv\frac{1}{N}\sum_{j=1}^{N}e^{\imath\theta_{j}}.\label{eq:Kuramoto-mean-field}
\end{equation}
(Models that use the $h^{th}$ harmonic need to consider the mean
field defined using $e^{\imath h\theta_{j}}$.) By a trick of complex
algebra\cite{strogatz2000fromkuramoto}, the dynamics of an individual
oscillator can be exactly rewritten in terms of interactions with
the mean field. For the Kuramoto model, the interactions take the
form
\begin{equation}
\dot{\theta}_{i}=\omega_{i}+r\,K\,\sin\left(\psi-\theta_{i}\right).\label{eq:Kuramoto-mean-field-model}
\end{equation}
If one could predict the behavior of $r$ and $\psi$, one would essentially
solve the dynamics of the Kuramoto model and could predict the properties
of interest for the system. Ott and Antonsen provide a method for
performing such a calculation for the Kuramoto model under the assumption
of a continuous oscillator density \cite{ott2008lowdimensional}.
Unfortunately, there is no generic closed form solution for $r\left(t\right)$
or $\psi\left(t\right)$ for an arbitrary \emph{finite} population
behaving according to the Kuramoto model or other similar models.

\begin{figure}
\includegraphics[width=0.8\columnwidth]{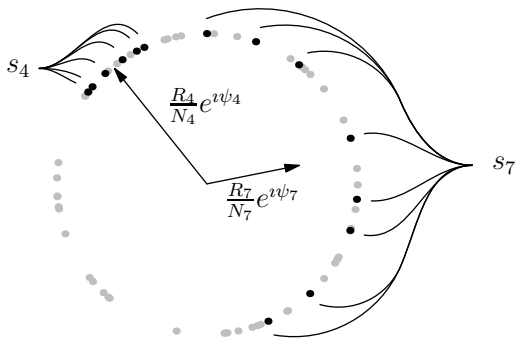}

\protect\caption{\label{fig:subsets-diagram}Depiction of two subsets (black dots)
in a full population (black and gray dots) of 64 oscillators, not
related to the simulation in figure \ref{fig:correlation-matrix}.
Subset 7 is more spread out than subset 4, and so has a smaller mean
field relative to subset size. If these subsets are coherent, their
constituent oscillators may get further apart or closer together,
but they do not lap each other.}
\end{figure}
How might we improve upon the mean-field reformulation for finite
sized systems? Consider the following minor refinement: split the
summation of the traditional order parameter into $\mathcal{N}$ pieces:
\begin{equation}
\sum_{j=1}^{N}e^{\imath\theta_{j}}=\sum_{j\in s_{1}}e^{\imath\theta_{j}}+\sum_{j\in s_{2}}e^{\imath\theta_{j}}+\cdots+\sum_{j\in s_{\mathcal{N}}}e^{\imath\theta_{j}}.
\end{equation}
In this expression, the $i^{th}$ oscillator's contribution $e^{\imath\theta_{i}}$
occurs only once on the left and right; that is, each oscillator is
assigned to only one subset, $s_{\ell}$. The assignment of oscillator
$\theta_{i}$ to subset $s_{\ell}$ is (for the moment) arbitrary
\footnote{Most of what I have to say about subset decomposition holds for any
random decomposition, and can even be generalized to fractional and
complex or imaginary membership.}. Taking a cue from the traditional analysis, define the \emph{subset}
\emph{mean field} for the $\ell^{th}$ subset as
\begin{align}
R_{\ell}e^{\imath\psi_{\ell}} & \equiv\sum_{j\in s_{\ell}}e^{\imath\theta_{j}}.\label{eq:subset-mean-field-definition}
\end{align}
Two such subsets, along with their mean fields, are depicted in figure
\ref{fig:subsets-diagram}. Employing the geometric interpretation
of complex numbers, we see that $\psi_{\ell}$ points roughly to their
average phase. If a subset's oscillators are near to each other then
$R_{\ell}\approx N_{\ell}$ and if they are scattered about the unit
circle then $R_{\ell}\ll N_{\ell}$. The subset's mean field serves
as a simple measure of the coherence and location of the subset, and
its formulation emerges from the original order parameter definition.

The traditional order parameter can be calculated, without any approximations,
in terms of these mean fields. A simple algebraic substitution leads
to
\begin{equation}
r\,e^{\imath\psi}=\frac{1}{N}\sum_{\ell=1}^{\mathcal{N}}R_{\ell}e^{\imath\psi_{\ell}}.\label{eq:full-op-in-terms-of-mean-fields}
\end{equation}
The behavior of each oscillator can also be written exactly in terms
of interactions with these mean fields. The specifics depend upon
the functions $P$ and $Q$ in equation \ref{eq:Winfree-model} or
$\Gamma$ in equation \ref{eq:Kuramoto-model-full-sum}, but such
a restatement can be obtained in terms of the harmonic mean fields
after employing the proper trigonometric identities. For example,
the oscillator dynamics for the Kuramoto model are 
\begin{equation}
\dot{\theta}_{i}=\omega_{i}+\frac{K}{N}\sum_{\ell=1}^{\mathcal{N}}R_{\ell}\sin\left(\psi_{\ell}-\theta_{i}\right),\label{eq:individual-in-terms-of-mean-fields-naive}
\end{equation}
which closely resembles the mean-field interaction given in equation
\ref{eq:Kuramoto-mean-field-model}. Again, this equation does not
rely on any approximations. The only limitation is that I have yet
to predict $R_{\ell}$ and $\psi_{\ell}$.

In order to predict the behavior of the subset mean fields, we need
an expression for their dynamics. Evaluating the time derivative of
the definition (equation \ref{eq:subset-mean-field-definition}) and
rearranging leads to
\begin{align}
\dot{R}_{\ell}+\imath\dot{\psi}_{\ell}R_{\ell} & =\sum_{j\in s_{\ell}}\imath\dot{\theta}_{j}e^{\imath\left(\theta_{j}-\psi_{\ell}\right)}\\
 & =\sum_{j\in s_{\ell}}\dot{\theta}_{j}\sin\left(\psi_{\ell}-\theta_{j}\right)+\imath\sum_{j\in s_{\ell}}\dot{\theta}_{j}\cos\left(\psi_{\ell}-\theta_{j}\right).\label{eq:complex-subset-dynamics}
\end{align}
The real and imaginary components give separate equations for the
evolution of a subset amplitude and phase, respectively. To obtain
the equations for a specific model, $\dot{\theta}_{j}$ must be replaced
with the expression for the time derivative specific to that model.
Although the dynamics in $\dot{\theta}_{j}$ depend upon all of the
mean fields, judicious use of trigonometric identities can lead to
separable contributions for internal dynamics and interactions between
subset mean fields. The internal dynamics are of particular interest:
if these could be approximated in terms of the subset's mean field,
it would be possible to coarse-grain the dynamics of the oscillator
model into a model for interacting amplitude oscillators representing
the subsets. Such a coarse graining relies upon a good approximation
for $\psi_{\ell}-\theta_{j}$.

\paragraph{Coherent subset approximation}

One good approximation for $\psi_{\ell}-\theta_{j}$ that is independent
of model details is the \emph{coherent subset approximation}. The
approximation is built on two observations about the clumps discussed
on the first page:
\begin{enumerate}
\item Natural ordering: For oscillators in a coherent subset, the positions
$\theta_{i}$ order according to increasing natural speed, $\omega_{i}$.
\item Uniform tidal effects: Relative oscillator positions fluctuate \emph{nearly
in unison}. For oscillators in a coherent subset $\ell$, relative
positions can be related to the subset's mean field amplitude, $R_{\ell}$.
\end{enumerate}
While uniform tidal effects have not previously been noted, evidence
of natural ordering is abundant in the literature, beginning with
Winfree's discussion of syntalansis in his original paper on the topic
\cite{winfree1967biological}. One mathematical approximation that
reflects these two observations is 
\begin{equation}
\theta_{j}-\psi_{\ell}\approx C\left(\omega_{j}-\bar{\omega}_{\ell}\right)^{\alpha}\left(N_{\ell}-R_{\ell}\right)^{\beta}.\label{eq:coherent-subpopulation-exponents}
\end{equation}
The $\omega$-dependence reflects the observation of natural ordering.
The dependence on the subset's mean field amplitude and the neglect
of other time dependence reflects the observation of uniform tidal
effects.

In the simplest calculation, the constant $C$ and exponents $\alpha$
and $\beta$ \emph{do not depend} on the details of the phase oscillator
model, but only on the definition of the subset mean field. With a
small rearrangement, equation \ref{eq:subset-mean-field-definition}
becomes 
\begin{equation}
R_{\ell}=\sum_{j\in s_{\ell}}e^{\imath\left(\theta_{j}-\psi_{\ell}\right)}.\label{eq:subset-mean-field-rearranged}
\end{equation}
To obtain constraints on the constant and exponents, expand the real
and imaginary components of equation \ref{eq:subset-mean-field-rearranged}
to second order in $\theta_{j}-\psi_{\ell}$. The imaginary component
gives
\begin{align}
0 & =\sum_{j\in s_{\ell}}\sin\left(\theta_{j}-\psi_{\ell}\right)\nonumber \\
 & \approx C\left(N_{\ell}-R_{\ell}\right)^{\beta}\sum_{j\in s_{\ell}}\left(\omega_{j}-\bar{\omega}_{\ell}\right)^{\alpha},
\end{align}
which can only hold generically if $\alpha\equiv1$. The real component
gives
\begin{align}
R_{\ell} & =\sum_{j\in s_{\ell}}\cos\left(\theta_{j}-\psi_{\ell}\right)\nonumber \\
 & \approx N_{\ell}-\frac{C^{2}}{2}\left(N_{\ell}-R_{\ell}\right)^{2\beta}\sum_{j\in s_{\ell}}\left(\omega_{j}-\bar{\omega}_{\ell}\right)^{2\alpha},
\end{align}
which implies $\beta=1/2$ and $C^{-2}=\frac{1}{2}\sum_{j\in s_{\ell}}\left(\omega_{j}-\bar{\omega}_{\ell}\right)^{2}\equiv\Delta_{\ell}^{2}$.
Finally we arrive at

\begin{equation}
\theta_{j}-\psi_{\ell}\approx\frac{\omega_{j}-\bar{\omega}_{\ell}}{\Delta_{\ell}}\sqrt{N_{\ell}-R_{\ell}},\label{eq:entrained-subset-approximation}
\end{equation}
This equation is the key result of this letter. Notice that the derivation
does not rely on a particular form of $P$, $Q$, or $\Gamma$. Although
this equation does not predict which oscillators will aggregate into
coherent subsets, it gives a model-independent time-dependent prediction
for the steady-state structure of such subsets.

\begin{figure}
\includegraphics[width=0.8\columnwidth]{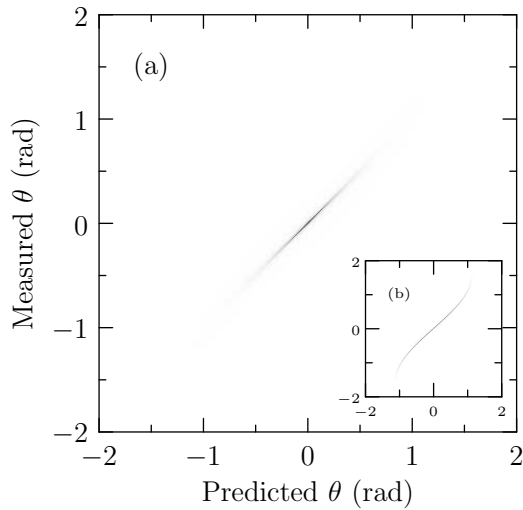}

\protect\caption{\label{fig:Relative-phase-positions}Histogram of actual (y) vs predicted
(x) relative phase positions for the approximation given by equations
\ref{eq:entrained-subset-approximation} for the Kuramoto model. Intensity
is linear in density. Figure (a) shows results for small subsets,
each of which constitute fewer than 10\% of their simulation's population.
Figure (b) shows results for large subsets, each of which have more
than 10\% of their simulation's population. The data represent aggregated
results of many simulations with a variety of coupling strengths and
full population sizes, and with emergent subsets of various sizes.
The correlation between the prediction and the measurement is $r^{2}=0.86$
for (a), $r^{2}=0.99$ for (b).}
\end{figure}

\paragraph{Agreement with simulation}

The agreement between the actual relative phases for the Kuramoto
model and the predictions of equation \ref{eq:entrained-subset-approximation}
is shown in the density plots given in figure \ref{fig:Relative-phase-positions}
and its inset. The main figure represents data from small subsets
while the inset represents data from large subsets. The figure includes
data from many different empirically identified subsets, themselves
constituents of about 80 unrelated unimodal populations of various
sizes, and all simulated with five coupling strengths. The coupling
strengths are chosen randomly between 0 and 2; the critical coupling
strength for these simulations is roughly $1.6$ \cite{strogatz2000fromkuramoto}.
Additional details as well as other phase oscillator models are discussed
in the supplementary material.

Figure \ref{fig:Relative-phase-positions} illustrates a strong correlation
between the measurement and prediction: the approximation works well.
Other models, such as the Ariaratnam-Strogatz model \cite{ariaratnam2001phasediagram}
or a third harmonic approximation to the sawtooth function, exhibit
equally impressive small-subset results. The prediction is not perfect,
with nonlinearities particularly prominent in the inset at large relative
positions. It is perhaps unsurprising that large-subset data exhibit
model-specific nonlinearities: these agree with the predictions of
Sakaguchi and Kuramoto \cite{sakaguchi1986asoluble}. It is remarkable,
however, that the nonlinearities are one-to-one. This one-to-one nature
strongly suggests that more accurate predictions can be expressed
as model-specific power series of equation \ref{eq:entrained-subset-approximation}.
In other words, for models that exhibit natural ordering and uniform
tidal effects, equation \ref{eq:entrained-subset-approximation} serves
as a universal starting point.

\paragraph{Discussion and conclusion}

The coherent subset approximation differs from similar results in
the literature in key ways. Sakaguchi and Kuramoto predicted entrained
relative phases for their model in ref \cite{sakaguchi1986asoluble}
equation 5a. Their predictions only applied to oscillators entrained
to the order parameter whereas equation \ref{eq:entrained-subset-approximation}
describes all coherent behavior, whether the mean field is appreciable
or negligible, and whether or not the oscillators of interest are
entrained to the mean field. It also gives a prediction for the instantaneous
relative position which does not rely on long-time averaging or the
coupling strength. Others have predicted the oscillator density distribution,
$\rho\left(\theta,\omega,t\right)$, for the Kuramoto model with infinite
system size and various population distributions \cite{acebron2005thekuramoto,ott2008lowdimensional,iatsenko2013stationary}.
Equation \ref{eq:entrained-subset-approximation} is distinct from
those predictions by addressing finite-size synchronization for a
broad set of models, and by imposing no restriction on the population
distribution. The coherent subset approximation is both more precise
and more general than previous predictions.

The coherent subset approximation provides a new angle for analyzing
distributions with finite support. In their impressive analysis, Martens
et al. analytically compute nearly all of the bifurcation diagram
for the Kuramoto model with a Lorentzian bimodal frequency distribution
\cite{martens2009exactresults}. As explained in the supplementary
material, obtaining the saddle-node infinite-period (SNIPER) bifurcation
curve for small $\tilde{\sigma}$ is simple using the coherent subset
approximation, and leads to unexpected scaling. Martens predicted
that $\tilde{\sigma}\propto2-\tilde{\omega}_{0}$ as $\tilde{\omega}_{0}\to2$.
The coherent subset approximation, on the other hand, predicts that
$\tilde{\sigma}\propto\sqrt{2-\tilde{\omega}_{0}}$ as $\tilde{\omega}_{0}\to2$.
The range of values over which the scaling takes this form is not
yet clear, but it should hold for any sufficiently narrow distribution
with finite support. Whether an infinite bimodal Gaussian distribution
would have linear or square-root scaling is intriguing but unknown.

All globally coupled phase oscillator models can be decomposed. The
evolution of the subset mean fields is given by equation \ref{eq:complex-subset-dynamics},
even those models which do not exhibit natural ordering and uniform
tidal effects. Unfortunately, the mean field dynamics given by equation
\ref{eq:complex-subset-dynamics} depend on model-specific details
and so do not coarse grain to a universal form. If universality exists
across a broad set of phase oscillator models, it will arise because
the statistics of the aggregations do not depend upon the underlying
model. This work provides the basis for such an analysis, but any
such claims go beyond the scope of this letter.

In this letter, I have shown that the dynamics of globally coupled
phase oscillator models can be restated exactly in terms of interactions
among and within subsets. Phase oscillator models that exhibit natural
ordering and uniform tidal effects will exhibit coherent subsets they
have nearly identical microstructure. The approximation given in equation
\ref{eq:entrained-subset-approximation} appears to perform well and
could serve as a starting point for more accurate approximations in
a wide variety of models. Rather than relying on long-time or ensemble
averaging to make meaningful predictions, an analysis based on subsets
provides a formulation for predicting features of individual populations,
paving the way for insights into a decades old problem in dynamic
phase transitions.
\begin{acknowledgments}
I would like to thank Georgios Tsekenis for fruitful discussion and
insightful feedback during the preparation of this paper.
\end{acknowledgments}

\bibliographystyle{unsrt}
\bibliography{references}

\begin{thebibliography}{10}

\bibitem{strogatz2000fromkuramoto}
S.~Strogatz.
\newblock From {Kuramoto} to {Crawford}: exploring the onset of synchronization
  in populations of coupled oscillators.
\newblock {\em Physica D}, 143(1):1--20, 2000.

\bibitem{winfree1967biological}
Arthur~T. Winfree.
\newblock Biological rhythms and the behavior of populations of coupled
  oscillators.
\newblock {\em Journal of Theoretical Biology}, 16(1):15--42, July 1967.

\bibitem{kuramoto1975selfentrainment}
Y.~Kuramoto.
\newblock Self-entrainment of a population of coupled nonlinear oscillators.
\newblock volume~39 of {\em Lecture Notes in Physics}, page 420.
  Springer-Verlag, 1975.

\bibitem{ariaratnam2001phasediagram}
Joel Ariaratnam and Steven Strogatz.
\newblock Phase {Diagram} for the {Winfree} {Model} of {Coupled} {Nonlinear}
  {Oscillators}.
\newblock {\em Physical Review Letters}, 86(19):4278--4281, May 2001.

\bibitem{sakaguchi1986asoluble}
Hidetsugu Sakaguchi and Yoshiki Kuramoto.
\newblock A soluble active rotator model showing phase transitions via mutual
  entrainment.
\newblock {\em Prog. Theor. Phys.}, 76(3):576--581, September 1986.

\bibitem{weisenfeld1998frequency}
Kurt Weisenfeld, Pere Colet, and Steven~H. Strogatz.
\newblock Frequency locking in {Josephson} arrays: {Connection} with the
  {Kuramoto} model.
\newblock {\em Phys. Rev. E}, 57(2):1563--1569, February 1998.

\bibitem{kourtchatov1995theoryof}
S.~Yu Kourtchatov, V.~V. Likhanskii, A.~P. Napartovich, F.~T. Arecchi, and
  A.~Lapucci.
\newblock Theory of phase locking of globally coupled laser arrays.
\newblock {\em Phys. Rev. A}, 52(5):4089--4094, November 1995.

\bibitem{temirbayev2012experiments}
Amirkhan~A. Temirbayev, Zeinulla~Zh. Zhanabaev, Stanislav~B. Tarasov,
  Vladimir~I. Ponomarenko, and Michael Rosenblum.
\newblock Experiments on oscillator ensembles with global nonlinear coupling.
\newblock {\em Physical Review E}, 85(1):015204, January 2012.

\bibitem{kiss2002emerging}
Istv{\'a}n~Z. Kiss, Yumei Zhai, and John~L. Hudson.
\newblock Emerging {Coherence} in a {Population} of {Chemical} {Oscillators}.
\newblock {\em Science}, 296(5573):1676--1678, April 2002.

\bibitem{kiss2008resonance}
Istv{\'a}n~Z. Kiss.
\newblock Resonance clustering in globally coupled electrochemical oscillators
  with external forcing.
\newblock {\em Physical Review E}, 77(4), April 2008.

\bibitem{mertens2011synchronization}
David Mertens and Richard Weaver.
\newblock Synchronization and stimulated emission in an array of mechanical
  phase oscillators on a resonant support.
\newblock {\em Physical Review E}, 83(4):046221, April 2011.

\bibitem{mertens2011individual}
David Mertens and Richard Weaver.
\newblock Individual and collective behavior of vibrating motors interacting
  through a resonant plate.
\newblock {\em Complexity}, 16(5):45--53, 2011.

\bibitem{acebron2005thekuramoto}
Juan~A. Acebr{\'o}n, Luis~L{\'o}pez Bonilla, Conrad J.~P{\'e}rez Vicente,
  F{\'e}lix Ritort, and Renato Spigler.
\newblock The kuramoto model: A simple paradigm for synchronization phenomena.
\newblock {\em Reviews of modern physics}, 77(1):137, 2005.

\bibitem{ott2008lowdimensional}
Edward Ott and Thomas~M. Antonsen.
\newblock Low dimensional behavior of large systems of globally coupled
  oscillators.
\newblock {\em Chaos: An Interdisciplinary Journal of Nonlinear Science},
  18(3):037113, 2008.

\bibitem{daido1989intrinsic}
H.~Daido.
\newblock Intrinsic {Fluctuation} and {Its} {Critical} {Scaling} in a {Class}
  of {Populations} of {Oscillators} with {Distributed} {Frequencies}.
\newblock {\em Prog. Theor. Phys.}, 81(4):727--731, April 1989.

\bibitem{chulho2013extended}
Choi Chulho, Ha~Meesoon, and Byungnam Kahng.
\newblock Extended finite-size scaling of synchronized coupled oscillators.
\newblock {\em Phys. Rev. E}, 88:032126--1--032126--7, September 2013.

\bibitem{buice2007correlations}
Michael~A. Buice and Carson~C. Chow.
\newblock Correlations, fluctuations, and stability of a finite-size network of
  coupled oscillators.
\newblock {\em Physical Review E}, 76(3):031118, 2007.

\bibitem{maistrenko2004mechanism}
Yu. Maistrenko, O.~Popovych, O.~Burylko, and P.~A. Tass.
\newblock Mechanism of {Desynchronization} in the {Finite}-{Dimensional}
  {Kuramoto} {Model}.
\newblock {\em Physical Review Letters}, 93(8):084102, August 2004.

\bibitem{iatsenko2013stationary}
D.~Iatsenko, S.~Petkoski, P.~V.~E. McClintock, and A.~Stefanovska.
\newblock Stationary and {Traveling} {Wave} {States} of the {Kuramoto} {Model}
  with an {Arbitrary} {Distribution} of {Frequencies} and {Coupling}
  {Strengths}.
\newblock {\em Physical Review Letters}, 110(6):064101, February 2013.

\bibitem{martens2009exactresults}
E.~Martens, E.~Barreto, S.~Strogatz, E.~Ott, P.~So, and T.~Antonsen.
\newblock Exact results for the {Kuramoto} model with a bimodal frequency
  distribution.
\newblock {\em Physical Review E}, 79(2), February 2009.

\end{thebibliography}


\begin{thebibliography}{1}

\bibitem{chulho2013extended}
Choi Chulho, Ha~Meesoon, and Byungnam Kahng.
\newblock Extended finite-size scaling of synchronized coupled oscillators.
\newblock {\em Phys. Rev. E}, 88:032126--1--032126--7, September 2013.

\bibitem{white2000acomparison}
Jr. White, K.P., M.J. Cobb, and S.C. Spratt.
\newblock A comparison of five steady-state truncation heuristics for
  simulation.
\newblock In {\em Simulation {Conference}, 2000. {Proceedings}. {Winter}},
  volume~1, pages 755--760 vol.1, 2000.

\bibitem{ariaratnam2001phasediagram}
Joel Ariaratnam and Steven Strogatz.
\newblock Phase {Diagram} for the {Winfree} {Model} of {Coupled} {Nonlinear}
  {Oscillators}.
\newblock {\em Physical Review Letters}, 86(19):4278--4281, May 2001.

\bibitem{martens2009exactresults}
E.~Martens, E.~Barreto, S.~Strogatz, E.~Ott, P.~So, and T.~Antonsen.
\newblock Exact results for the {Kuramoto} model with a bimodal frequency
  distribution.
\newblock {\em Physical Review E}, 79(2), February 2009.

\end{thebibliography}

\end{document}

\section{Numerical Setup}

For all simulations, the population is sampled from a normal distribution
of width $\sigma=1\nicefrac{rad}{s}$ and mean $\bar{\omega}=1\nicefrac{rad}{s}$.
The mean is inconsequential for the Kuramoto model but matters for
one of the models considered below. In all simulations, I choose positions
that are randomly distributed around the unit circle. The number of
oscillators for a given simulation is sampled randomly from the uniform
distribution between 30 and 5000.

I choose $K$ values randomly from the uniform distribution between
0 and 2. Other models use other ranges as detailed below. The coupling
ranges are based on trial and error, aiming to get a good sampling
of unsynchronized and synchronized behavior for each model.

In all simulations presented, I follow the well-established practice
of using a fourth-order Runge-Kutta method with a time step of 0.01s
\cite{chulho2013extended}. Based on the numerical evidence of Chulho
et al. \cite{chulho2013extended}, the transient behavior of the Kuramoto
model lasts for about $3\sqrt{N_{osc}}$ seconds. I run all simulations
for $900\sqrt{N_{osc}}$ seconds, i.e. 30 multiples of the transient
duration, storing the order parameter and oscillator positions each
second, i.e. every 100 time steps. I then use the Mean Squared Error
Rule to identify when the order parameter reaches steady state \cite{white2000acomparison},
using either that or $3\sqrt{N_{osc}}$ seconds, whichever is larger.

\section{Identifying Entrained Subsets of Oscillators}

In addition to the order parameter and oscillator positions, I also
monitor the phase differences between oscillators with consecutive
natural speeds, and make note of all $2\pi$ phase slips and the time
step at which they occur. To identify coherent subsets, discard all
phase slips that occur before the onset of the steady state and identify
all pairs with \emph{no} slips for the duration of the steady state.
Such pairs are assumed to be mutually entrained. If one oscillator
is entrained to two other oscillators, then all three of them must
be mutually entrained. Using this approach, I extract the entrained
subsets empirically.

\section{Producing the Third Figure}

To produce the third figure in the letter, I select 1000 random moments
in time, $t_{n}$, during the steady-state of each subset. I compute
the subset mean field, $R_{\ell}\left(t_{n}\right)e^{\imath\psi_{\ell}\left(t_{n}\right)}$,
from the $\theta_{j}\left(t_{n}\right)$. I then compute the actual
relative position, $\theta_{j}\left(t_{n}\right)-\psi_{\ell}\left(t_{n}\right)$,
as well as the prediction for the relative positions based on $R_{\ell}\left(t_{n}\right)$.
The plots are two-dimensional histograms depicting the number of times
a predicted value of $\theta_{j}-\psi_{\ell}$ for any one oscillator
corresponded with any measured value of $\theta_{j}-\psi_{\ell}$.

\section{Additional Models}

\begin{figure}
\includegraphics[width=0.98\columnwidth]{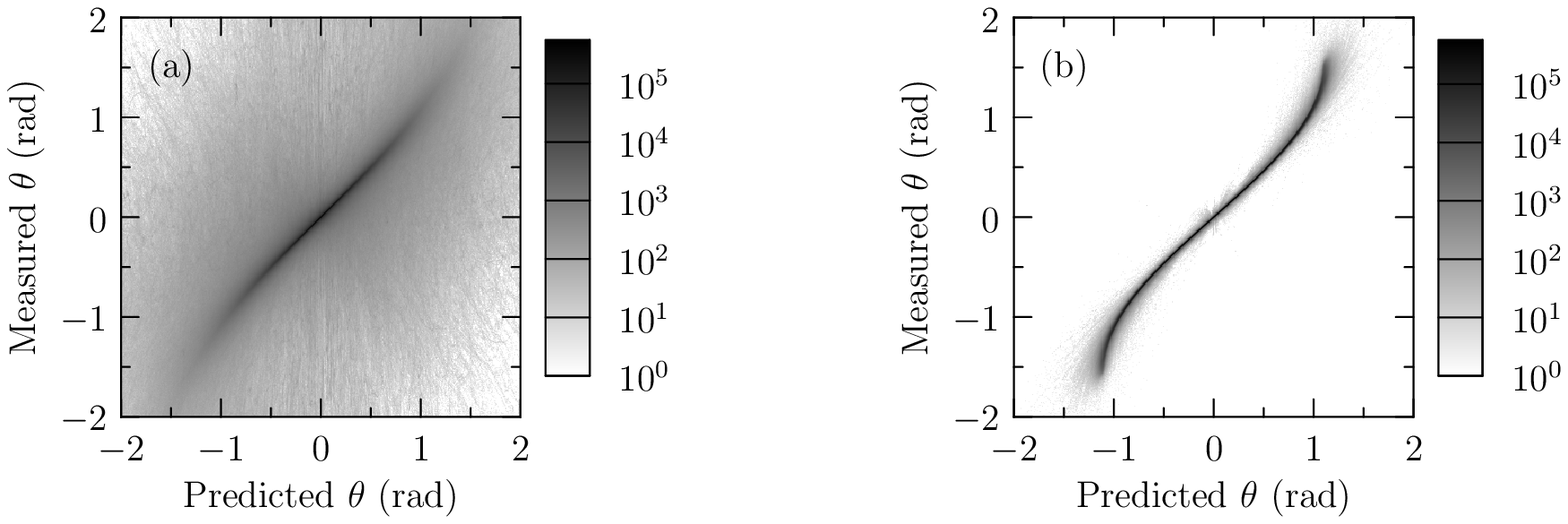}

\protect\caption{\label{fig:Kuramoto}Prediction vs simulation for the Kuramoto model.
Figure (a) is for small coherent subsets while figure (b) is for large
coherent subsets. In contrast to the figure in the letter, the intensity
indicates the logarithm of the density, rather than the density itself.}
\end{figure}
\begin{figure}
\includegraphics[width=0.98\columnwidth]{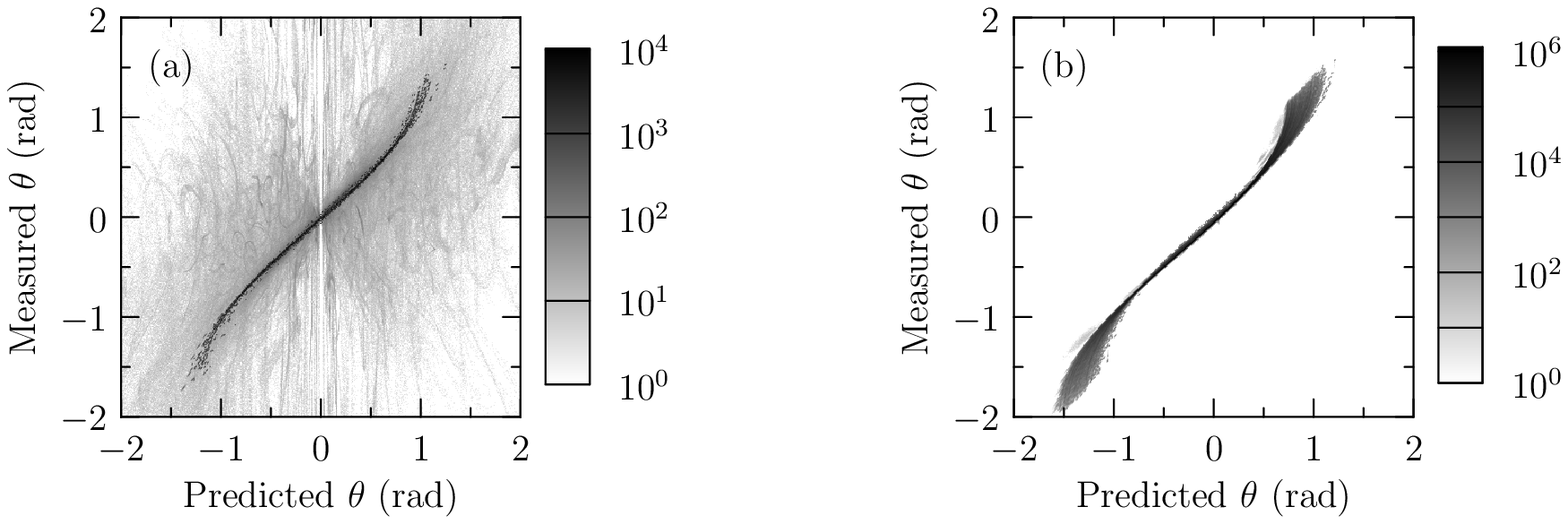}

\protect\caption{\label{fig:Ariaratnam}Prediction vs simulation for the Ariaratnam-Strogatz
model, a Winfree model. Figure (a) is for small coherent subsets while
figure (b) is for large coherent subsets. The correlation for the
small subsets is $r^{2}=0.99$ and for the large subsets is $r^{2}=0.99$.}
\end{figure}
\begin{figure}
\includegraphics[width=0.98\columnwidth]{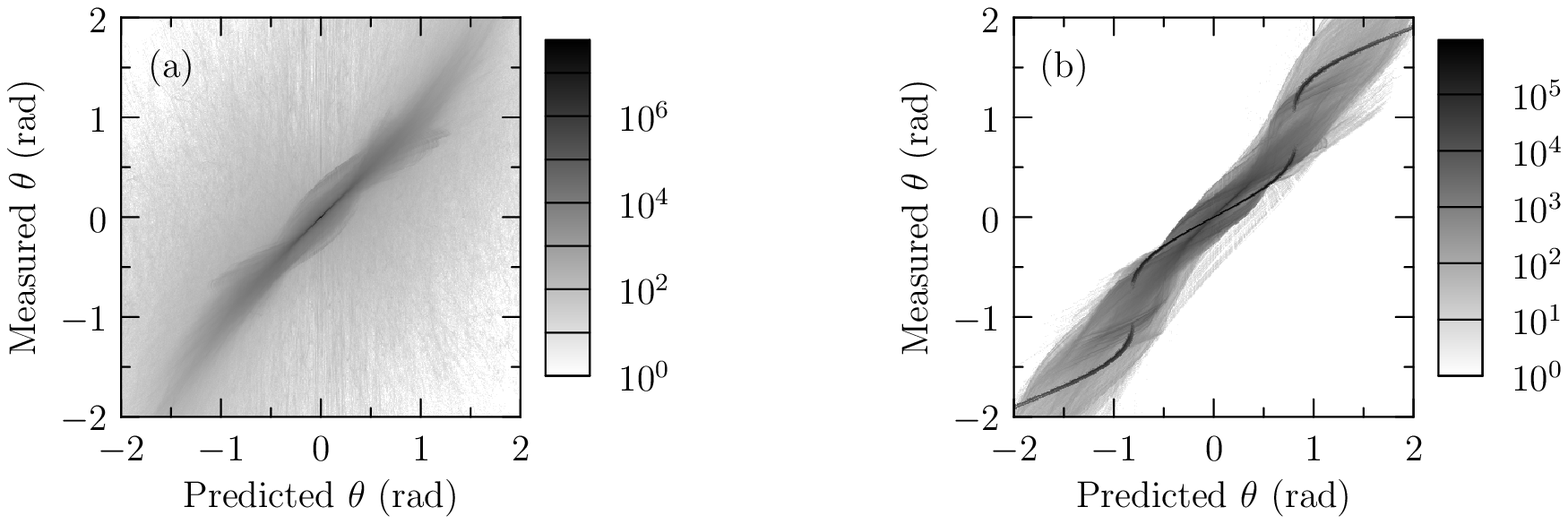}

\protect\caption{\label{fig:sawtooth}Prediction vs simulation for the third-harmonic
sawtooth model, a Kuramoto-like model. Figure (a) is for small coherent
subsets while figure (b) is for large coherent subsets. The correlation
for the main figure is $r^{2}=0.86$ and for the inset is $r^{2}=0.95$.}
\end{figure}
In this section I present the results of the Kuramoto model alongside
the results of two other globally coupled phase oscillator models.

Figure \ref{fig:Kuramoto} shows the small and large histograms for
the Kuramoto model, using the same data as in the letter. In contrast
to the figures in the main letter, the intensity indicates the logarithm
of the density. This is necessary to represent the results of the
sawtooth model, discussed below. Even with heightened sensitivity
to small densities, the interpretation remains the same: relative
positions are well approximated by the coherent subset approximation
for small subsets, and one-to-one nonlinearities arise for large subsets.

Ariaratnam and Strogatz \cite{ariaratnam2001phasediagram} considered
a Winfree model for which $P\left(\theta_{j}\right)=-1-\cos\theta_{j}$
and $Q\left(\theta_{i}\right)=\sin\theta_{i}$. Figure \ref{fig:Ariaratnam}
shows the small and large subset histograms for the model. For these
simulations, the coupling was chosen randomly from the uniform distribution
between 0 and 1. Strong correlations and nonlinear effects occur for
subsets greater than 3\% of the full population size, substantially
smaller than for the Kuramoto model.

Note that the mean frequency is significant in this model. This model
can be interpreted as a Kuramoto-like model in which the oscillators
feel a pull between a ``uniform field'' directed toward $\theta=0$
coupled with strength $K$, a mean field coupled with strength $K/2$,
and an odd additive term---$\sin\left(\theta_{i}+\theta_{j}\right)$---also
coupled with strength $K/2$. We expect the mean field to have two
elements: a component aligned to the uniform field, and a component
rotating in spite of the uniform field. Large $\bar{\omega}$ should
lead to more oscillators in the rotating component. Ariaratnam and
Strogatz considered an evenly spaced distribution of oscillators,
without disorder, so it is not known exactly how the choice of $\bar{\omega}$
will effect the results, but an effect is expected.

The final model is a Kuramoto-like model based on the Fourier expansion
of the sawtooth function truncated at the third harmonic, for which
$\Gamma\left(\Delta\theta\right)=\sin\Delta\theta-\frac{1}{2}\sin2\Delta\theta+\frac{1}{3}\sin3\Delta\theta$.
The results for this model are shown in figure \ref{fig:sawtooth}.
The coupling was chosen randomly from the uniform distribution between
0 and 4. The cutoff for large subsets was empirically set at 45\%,
substantially larger than for the Kuramoto model. It is also notable
that the vast majority of the small-subset density lies within very
small relative angles.

As mentioned in the main text, the predictions work fairly well, and
although nonlinearities are present, the predictions still seem to
have a one-to-one mapping to the measurements for most data. It is
particularly interesting to note that large subset nonlinearities
for the Kuramoto and the third-order sawtooth closely resemble the
actual form of the coupling functions themselves!

\section{Square-root Scaling of the SNIPER Bifurcation Curve}

In this section I will obtain a finite-size correction to the phase
diagram of Martens et al \cite{martens2009exactresults}. To begin,
I insert the form of $\dot{\theta}_{j}$ from the Kuramoto model into
equation 11 in the letter. I also substitute the coherent subset approximation,
equation 16 in the letter, and expand the trigonometric sums to second
order to obtain
\begin{align}
\dot{R}_{\ell} & =-2\,\Delta_{\ell}\sqrt{N_{\ell}-R_{\ell}}+2\left(N_{\ell}-R_{\ell}\right)\frac{K}{N}\sum_{m}R_{m}\cos\left(\psi_{m}-\psi_{\ell}\right),\\
\dot{\psi}_{\ell}R_{\ell} & =\bar{\omega}_{\ell}R_{\ell}+\left(2\,R_{\ell}-N_{\ell}\right)\frac{K}{N}\sum_{m}R_{m}\sin\left(\psi_{m}-\psi_{\ell}\right).
\end{align}
For a symmetric bimodal population, for which all the oscillators
are members of one of two coherent subsets $s_{1}$ or $s_{2}$, I
can obtain an expression for the mean field amplitudes $R_{1}=R_{2}\equiv R_{\ell}$
as well as the difference of the phases, $\Delta\psi\equiv\psi_{2}-\psi_{2}$:
\begin{align}
\dot{R}_{\ell} & =-2\Delta_{\ell}\sqrt{N_{\ell}-R_{\ell}}+2\left(N_{\ell}-R_{\ell}\right)\frac{K}{N}R_{\ell}\left(1+\cos\Delta\psi\right),\\
\frac{d}{dt}\Delta\psi & =\bar{\omega}_{2}-\bar{\omega}_{1}-2\frac{2\,R_{\ell}-N_{\ell}}{N}K\sin\Delta\psi.
\end{align}

Now suppose that the peaks of the bimodal distribution are very narrow
with respect to the coupling strength and their separation. In that
case, the oscillators belonging to each peak should mutually entrain,
behaving like two giant oscillators. The saddle-node infinite-period
(SNIPER) bifurcation separates the the phase space where these two
giant oscillators are locked together or are counter-propagating.
The dynamics for the locked oscillators are simply the fixed points
given by $\dot{R}_{\ell}=0$ and $\frac{d}{dt}\Delta\psi=0$, so
\begin{align}
2\Delta_{\ell}\sqrt{N_{\ell}-R_{\ell}} & =2\left(N_{\ell}-R_{\ell}\right)\frac{K}{N}R_{\ell}\left(1+\cos\Delta\psi\right),\\
\bar{\omega}_{2}-\bar{\omega}_{1} & =2\frac{2\,R_{\ell}-N_{\ell}}{N}K\sin\Delta\psi.
\end{align}
The saddle-node boundary occurs at values of $\Delta_{\ell}$ and
$\bar{\omega}_{2}-\bar{\omega}_{1}$ at which these equations can
not be satisfied. The most extreme separation between $\bar{\omega}_{2}-\bar{\omega}_{1}$
occurs when $\sin\Delta\psi\approx1$, in which case $\cos\Delta\psi\approx0$,
leading to
\begin{align}
2\Delta_{\ell}\sqrt{N_{\ell}-R_{\ell}} & =2\left(N_{\ell}-R_{\ell}\right)\frac{K}{N}R_{\ell},\\
\bar{\omega}_{2}-\bar{\omega}_{1} & =2\frac{2\,R_{\ell}-N_{\ell}}{N}K.
\end{align}

To obtain expressions using the notation of Martens, use 
\begin{align}
\bar{\omega}_{2}-\bar{\omega}_{1} & =2\,\omega_{0},\\
\Delta_{\ell}^{2} & =\frac{N_{\ell}-1}{2}\sigma_{\ell}^{2}.
\end{align}
Substituting this into the above expressions and eliminating $R_{\ell}$
leads to
\begin{equation}
\frac{4\,\sigma_{\ell}}{K}=\frac{1}{4}\sqrt{2-\frac{4\,\omega_{0}}{K}}\left(2+\frac{4\,\omega_{0}}{K}\right)\sqrt{\frac{2\,N}{N-2}}.
\end{equation}
Taking the large $N$ limit so that $N/\left(N-2\right)\approx1$,
and using Martens' definitions,
\begin{align}
\tilde{\sigma} & =\frac{4\,\sigma_{\ell}}{K},\\
\tilde{\omega}_{0} & =\frac{4\,\omega_{0}}{K},
\end{align}
I obtain 
\begin{equation}
\tilde{\sigma}=\frac{\sqrt{2}}{4}\sqrt{2-\tilde{\omega}_{0}}\left(2+\tilde{\omega}_{0}\right).
\end{equation}
In contrast, in the vicinity of $\tilde{\omega}_{0}\approx2$, Martens'
eq 33 comes to
\begin{equation}
\tilde{\Delta}=\frac{4}{7}\left(2-\tilde{\omega}_{0}\right),
\end{equation}
where $\Delta$ represents the half-width-half-maximum of the individual
peaks and $\tilde{\Delta}\equiv4\,\Delta/K$.

I observed in the previous section that the coherent subset approximation
is not exact for large separations $\psi_{\ell}-\theta_{j}$. These
inconsistencies limit the range of applicability of my calculation,
but do not invalidate it. For subsets with sufficiently narrow $\tilde{\sigma}$,
all relative positions should fall within the well-fit range of the
prediction. The square-root dependence will certainly hold for these
distributions.

\bibliographystyle{unsrt}
\bibliography{references}